\documentclass[12pt]{article}
\usepackage{times}
\usepackage{graphicx} 
\usepackage{float} 
\usepackage{bm} 
\usepackage{cite}
\usepackage{xcolor}
\usepackage{lineno}
\title{Real-time shaping of entangled photons by classical control and feedback}

\author
    {Ohad Lib, Giora Hasson, Yaron Bromberg$^\ast$\\
\\
\normalsize{Racah Institute of Physics, The Hebrew University of Jerusalem, Jerusalem, 91904 Israel}\\
\\
\normalsize{$^\ast$To whom correspondence should be addressed; E-mail:  Yaron.Bromberg@mail.huji.ac.il.}
}

\date{}

\begin{document} 

\maketitle

\begin{abstract}
Quantum technologies hold great promise for revolutionizing photonic applications such as cryptography. Yet their implementation in real-world scenarios is held back, mostly due to sensitivity of quantum light to scattering. Recent developments in optimizing the shape of single photons introduce new ways to control quantum light. Nevertheless, shaping single photons in real-time remains a challenge due to the weak associated signals, which are too noisy for optimization processes. Here, we overcome this challenge and control scattering of entangled photons by shaping the classical laser beam that stimulates their creation. We discover that since the classical beam and the entangled photons follow the same path, the strong classical signal can be used for optimizing the weak quantum signal. We show that this approach can increase the length of free-space turbulent quantum links by up to two orders of magnitude, opening the door for employing wavefront shaping for quantum communications.   
\end{abstract}
\section*{Introduction}
Nearly a century after quantum mechanics revolutionized the way we understand nature, quantum resources such as superposition and entanglement are beginning to enter and transform technology\cite{Dowling2003, OBrien2009}. Many implementations of quantum technologies such as quantum communication\cite{lo2014secure,Liao2017,takenaka2017satellite} and quantum imaging\cite{brida2010experimental,Lemos2014,Tenne2018superres}, are based on photonic platforms that encode quantum bits (qubits) using single photons. One of the main challenges in such applications is the low flux of single photons that can be sent per communication channel or image pixel, resulting in extremely low capacities of these systems. A promising approach for boosting the capacity of quantum systems is to encode multi-level quantum bits (coined \textit{qudits}) using a single photon in a d-dimensional Hilbert space\cite{bechmann2000largeralphabets,cerf2002dlevel,erhard2018twisted}. To this end, photonic qudits have been implemented in the temporal\cite{temporalqudit2017}, spectral\cite{lukens2017frequency} and spatial\cite{Osuallivan-Hale2005,walborn2006quditsspatial} domains. The spatial domain is in particular attractive, since using spatial light modulators (SLMs) it is possible to arbitrary rotate qudits in the d-dimensional space, simply by shaping the spatial distribution of photons\cite{kagalwala2017qubitslm}. However, this also implies that the information carried by spatial qudits is extremely sensitive to scattering and aberrations, acting as random rotations on the qudits. For example, scattering of entangled photons that encode spatial qudits, scrambles their unique quantum correlations, resulting in a random grainy spatial correlation pattern, coined two-photon speckle \cite{Beenakker2009,Peeters2010}.  Scattering of photonic qudits is therefore a limiting factor in implementing photonic quantum technologies in real-world applications, such as ground-satellite quantum communication\cite{lo2014secure,Liao2017,takenaka2017satellite} or quantum imaging of biological samples\cite{jin2019genuine}. In the past few years, extensive research was devoted for protecting the information carried by spatial qudits, by encoding them in spatial modes that are immune to scattering and aberrations\cite{vallone2014oam,Mirhosseini2015,krenn2016twistedoam,sit2017oam}. Nevertheless, for significant scattering all spatial modes will eventually suffer from scrambling of the information carried by the photons\cite{paterson2005atmospheric,tyler2009influence}.
 
A promising approach for cancelling scattering of classical light is wavefront shaping. Over a decade after the pioneering work of Vellekoop and Mosk\cite{Vellekoop2007}, who focused classical light through scattering media using an SLM, a remarkable set of tools for controlling light in random media has been developed\cite{mosk2012controlling,horstmeyer2015guidestar,vellekoop2015feedback}. It is therefore appealing to adopt these tools to the quantum regime. Over the past few years several important developments towards spatial control of
quantum light have been reported \cite{Defienne2014,Defienne2016,Wolterink2016,HugoPRL,peng2018manipulation}. Nevertheless, for practical applications such as quantum technologies, real-time optimization and feedback must be performed, which is in particular challenging due to the inherently weak quantum signals. In most demonstrations to date, photon-pairs generated by spontaneous parametric down conversion (SPDC) were sent to an SLM that directly shaped their spatial distribution before hitting a scattering sample, and the optimization was performed in advance using an auxiliary classical laser at the same wavelength as the entangled photons \cite{Defienne2014,Defienne2016,Wolterink2016,HugoPRL}. In a few other demonstrations, the SLM modulated the bright laser beam that stimulates the SPDC process (coined \textit{pump beam}), and feedback was provided by the inherently weak quantum signal\cite{pugh2016towards,peng2018manipulation}. Thus, although the significant progress made, all demonstrations to date do not overcome the fundamental challenge of providing fast feedback in real-time, and thus cannot be implemented in real-world applications.

In this work, we demonstrate a novel method for compensating the scattering of entangled photons which allows, for the first time, real-time optimization of quantum correlations (Fig.\ref{fig:1}a). Our method consists of two main components: shaping and feedback. First, instead of shaping the entangled photons directly, we use the well known technique of pump shaping\cite{Monken1998,pugh2016towards,peng2018manipulation} to control the correlations of the entangled photons, thus avoiding additional loss to the quantum link. Second and most importantly, we utilize for the first time the intensity of the classical pump beam for the feedback on the entangled photons. Since we use the classical pump beam both for control and feedback, the well-established toolbox of classical wavefront shaping can be trivially extended to the quantum regime, opening the door for implementing wavefront shaping in quantum technologies.
We explain this striking result by showing,  both theoretically and experimentally, that when the pump beam scatters by the same random sample as the entangled photons, the spatial distribution of its intensity is identical to the spatial correlations of the entangled photons. Hence, any manipulation of the classical pump intensity has an identical effect on the entangled photons correlations. Since wavefront shaping of a classical bright beam is much faster and more efficient than shaping weak fluxes of entangled photons, we were able, for the first time, to demonstrate real-time wavefront correction of entangled photon scattered by a dynamically moving diffuser. Finally, we show using numerical simulations that our method allows significant improvement of free-space quantum links by compensating the effect of turbulence on the entangled photons.
\section*{Results}
To explain why scattering of two entangled photons corresponds to
scattering of a single pump photon at half the wavelength, we write the quantum state of the two photons  (coined \textit{signal} and \textit{idler} photons), in terms of their transverse wavevector components  $\mathbf{q_s}$ and  $\mathbf{q_i}$, $\\|{\psi}\rangle=\int d \mathbf{q_s}d\mathbf{q_i}\psi(\mathbf{q_s,q_i})a^\dagger(\mathbf{q_s})a^\dagger(\mathbf{q_i})|0\rangle$. Here $a^\dagger(\mathbf{q})$ is the creation operator of a photon with a transverse momentum $\mathbf{q}$, $|0\rangle$ is the vacuum state, and we assume the signal and idler photons have the same frequency and polarization. The two-photon amplitude $\psi(\mathbf{q_s,q_i})$ can be expressed in terms of the angular spectrum of the pump beam $v(\mathbf{q})$\cite{Monken1998},
\begin{equation}\label{1}
\psi(\mathbf{q_s,q_i}) = v\left(\mathbf{q_s+q_i}\right)\Phi\left(
\mathbf{q_s-q_i}\right),
\end{equation}
where $\Phi(\mathbf{q})\propto Sinc(\frac{L}{4k}\mathbf{q}^2) $ is the phase matching function of the SPDC crystal, $L$ is the crystal length and $k$ is the pump wavenumber inside the crystal. The number of inseparable modes in the superposition of the two-photon state, quantified by the Schmidt number $K$, is proportional to the ratio between the width of the phase matching function $\Phi(\mathbf{q})$ (determined by the crystal length $L$), and the width of the pump angular spectrum function $v(\mathbf{q})$ (determined by the width of the pump beam). In the so called thin-crystal regime, $K\gg1$, the two-photon state can be approximated by $\psi(\mathbf{q_s,q_i}) \propto v\left(\mathbf{q_s+q_i}\right)$\cite{Monken1998}, yielding an Einstein-Podolsky-Rosen (EPR) entangled state\cite{howell2004realization}. In this regime, we can precisely control the two-photon amplitude, by tailoring the angular spectrum of the pump beam or equivalently, by controlling its spatial profile, $W(\boldsymbol{\rho})$, at the input plane of the crystal.

We start by considering a thin diffuser placed right after the crystal, that can be modeled by a linear transformation on the creation operator $a^\dagger(\mathbf{q}) \rightarrow\int d \bm{\rho} a^\dagger(\bm{\rho})A_d(\bm{\rho})\exp(i\bm{\rho}\cdot\mathbf{q})$, where $A_d(\bm{\rho})$ is the amplitude transfer function of the thin diffuser and $\bm{\rho} $ is the transverse spatial coordinate. We note that if  the diffuser is not located right after the crystal, it can always be re-imaged on the crystal plane, as in conjugate-plane adaptive optics \cite{beckers1988isoplanatic}.  Experimentally, the two-photon quantum state is measured using two single photon detectors placed at the far-field of the crystal. The rate of coincidence events, i.e. detection of two photons simultaneously, one photon with transverse wavevector $\mathbf{q_s}$ and the other with transverse wavevector $\mathbf{q_i}$,  is given by $C(\mathbf{q_s},\mathbf{q_i})=\left|\langle{0}|a(\mathbf{q_i})a(\mathbf{q_s})|\psi\rangle\right|^2$,  yielding (see Supplementary)\cite{Walborn2010},
\begin{equation}\label{2}
C(\mathbf{q_s,q_i})\propto \left|\int d\bm{\rho}W(\bm{\rho})A_d^2(\bm{\rho})\exp(-i\bm{\rho}\cdot(\mathbf{q_s+q_i}))\right|^2
\end{equation}
For random diffusers, the spatial distribution of the coincidence pattern  $C(\mathbf{q_s,q_i})$ exhibits a random two-photon speckle pattern\cite{Beenakker2009,Peeters2010}. In the absence of loss, the right hand side of Eq. (\ref{2}) corresponds to the far-field intensity profile of the pump beam in the direction $\mathbf{q}=\mathbf{q_i}+\mathbf{q_s}$ (see Supplementary). Hence, by measuring the intensity profile of the pump beam, we get the spatial distribution of the two-photon state. Since this is true for any pump profile $W(\bm\rho)$, using an SLM we can optimize the pump profile to get a focused pump spot at the far-field of the diffuser, and the quantum correlations between the photons will become spatially localized, simultaneously. In case the diffuser is lossy, the pump intensity and the coincidence pattern may not be identical. Nevertheless, optimizing the pump profile will still localize the two-photon spatial correlations (See Supplementary). The fact that pump-optimization simultaneously localizes the two-photon correlations is a remarkable feature of spatially entangled photons. It offers orders of magnitude faster feedback than would have been possible with the inherently weak signal provided by the coincidence rate, allowing us to extend wavefront shaping to the quantum domain, by applying classical wavefront shaping to the bright pump beam. 

\begin{figure}[H]
\centering
\includegraphics[width=1\textwidth]{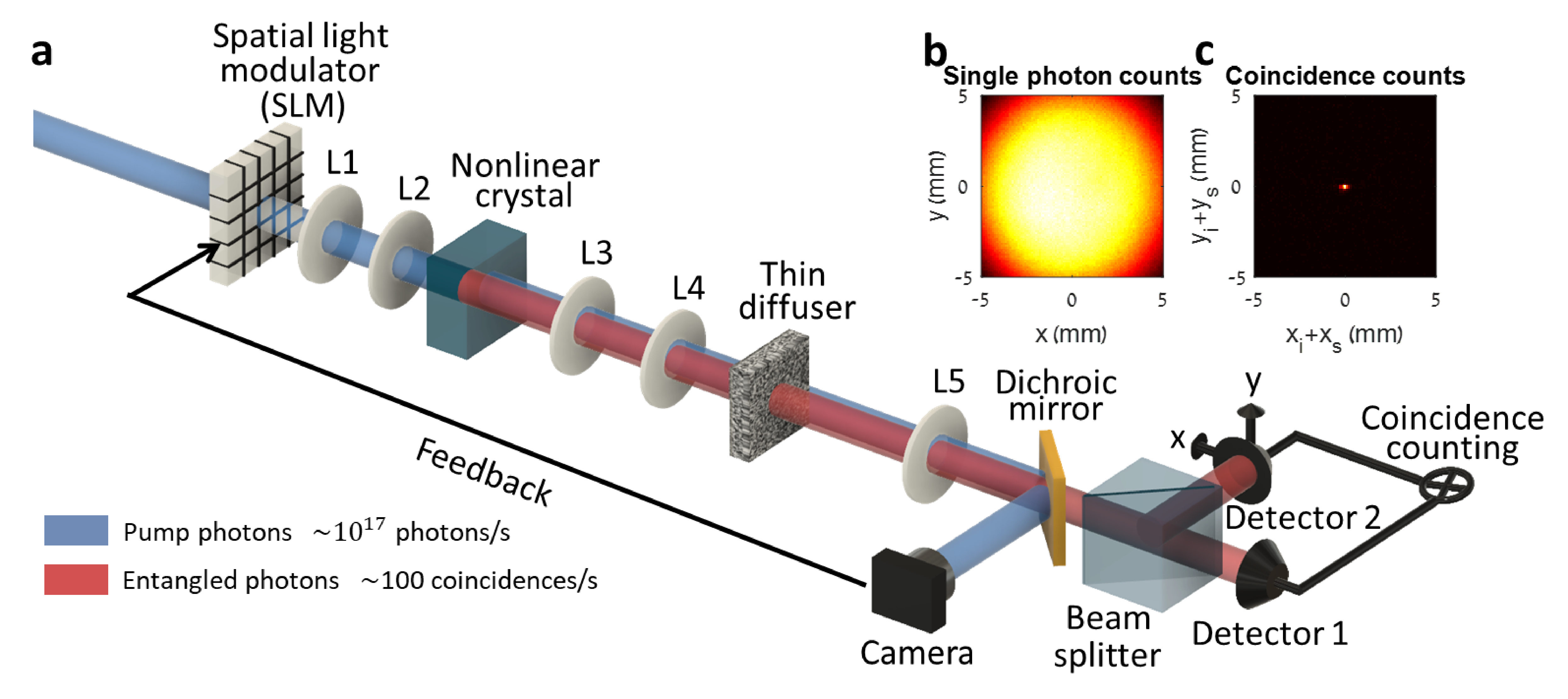}
  \caption{\label{fig:1}  Experimental setup. (a) Spatially entangled photons are created by pumping a nonlinear crystal (PPKTP) with a $\lambda=404nm$ continues-wave laser. Both the pump beam and the entangled photons pass through a thin diffuser, which is imaged on the crystal and SLM planes (by lenses $L1-L4$), and measured at the far-field (lens $L5$). An optimization method is employed for compensating the scattering of the pump beam using the SLM. (b) The far-field single counts, in the absence of the diffuser, is much wider than the coincidence distribution (c), indicating high spatial entanglement. }
\end{figure}

The quantum wavefront shaping experimental setup is depicted in Fig.\ref{fig:1}a. A $2mm$ long PPKTP crystal is pumped by a $\lambda=404nm$ continuous-wave laser, which is shaped by a phase-only SLM imaged on the input facet of the crystal. The SPDC process in the crystal generates a continuous flux of entangled photon-pairs, with a Schmidt number of  $K\approx680$ (Fig.\ref{fig:1}b,c, Supplementary). After the crystal, both the pump beam and the entangled photons are scattered by a thin diffuser located at the image plane of the crystal, creating a fully-developed speckle pattern with no ballistic component. The pump intensity and the two-photon coincidence patterns are measured at the far-field of the crystal, after separating the pump and entangled photons using a dichroic mirror. The coincidence patterns at the far-field are measured using two single photon detectors, where one detector is scanning while the other is always kept stationary.

\begin{figure}[H]
\centering
\includegraphics[width=0.75\textwidth]{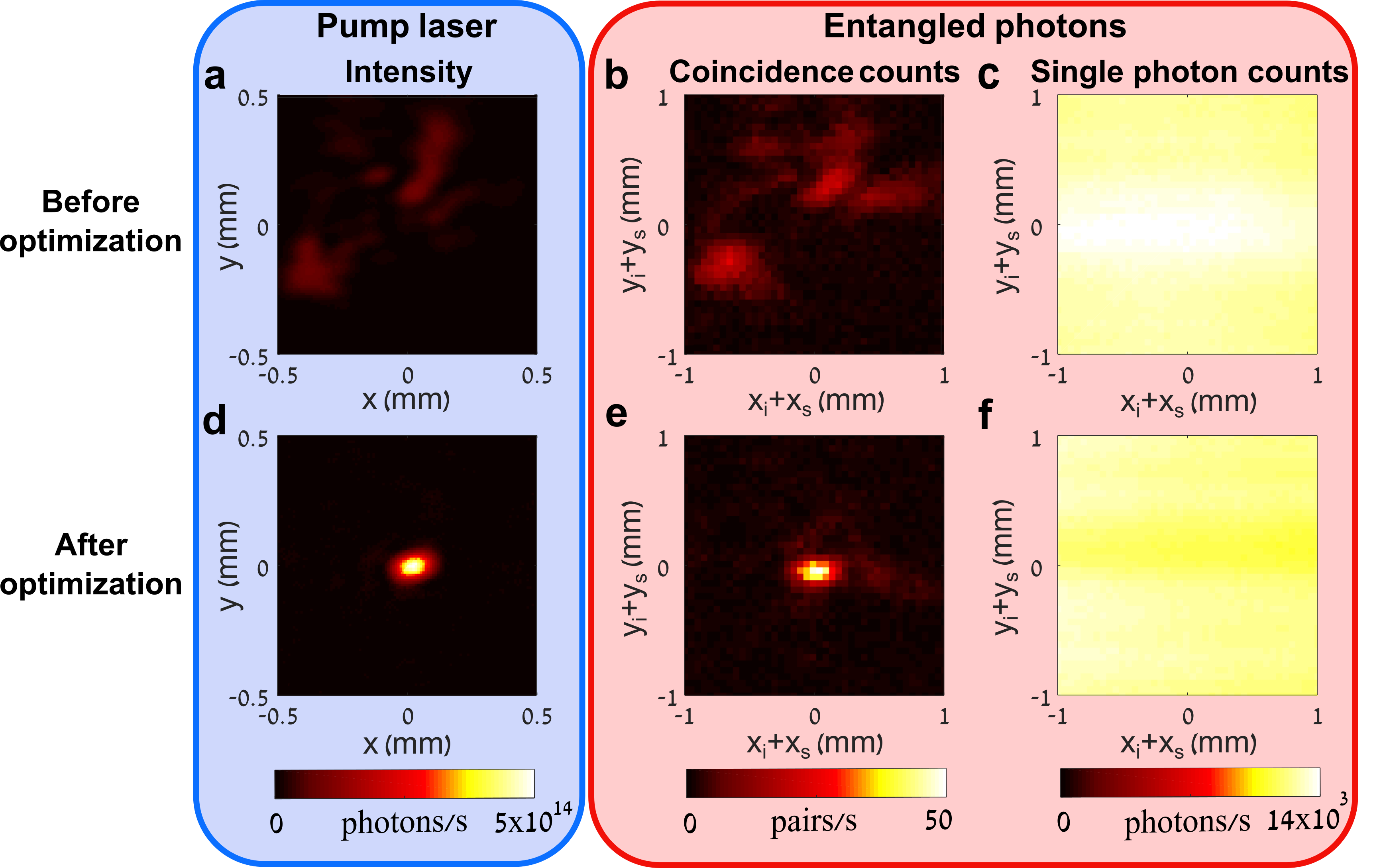}
\caption{\label{fig:2} Quantum wavefront shaping. The pump beam and the entangled photons pass through the same diffuser, forming similar speckle patterns in the intensity (a) and coincidence (b) pictures, respectively. By performing classical wavefront shaping on the bright pump beam, a single speckle grain is enhanced (d), yielding a simultaneous enhancement and localization of the quantum correlations at the corresponding location (e). The single photon counts (c,f) are not affected by either the scattering or the optimization.}
\end{figure}

Figures \ref{fig:2}a and \ref{fig:2}b depict the measured pump intensity and two-photon coincidence rate, at the far-field of the diffuser. Even though the wavelength of the entangled photons is twice the wavelength of the pump photons, the two signals exhibit strikingly similar patterns, as predicted by Eq. (\ref{2}). The wavelength difference comes into play only in the scaling of the two patterns, as the two-photon pattern is stretched by a factor of two compared to the pump pattern (see Supplementary). Since the pump and two-photon speckle patterns are remarkably similar, we can apply wavefront shaping optimization to the classical pump beam, and the quantum two-photon correlations will be optimized simultaneously. Specifically, we use the partitioning optimization algorithm to enhance the intensity of the pump beam at an arbitrary point at the far-field of the diffuser\cite{vellekoop2015feedback} (Fig.\ref{fig:2}d). With the exact same phase mask applied to the pump beam, the two-photon coincidence pattern is measured, showing a clear enhancement and localization of the two-photon correlations at the target area (Fig.\ref{fig:2}e). Interestingly, the single photon counts (Fig.\ref{fig:2}c,f) are not affected by the scattering nor by the optimization, due to the multimode nature of SPDC light in the high Schmidt number regime\cite{Peeters2010}. 

\begin{figure}[H]
\centering
\includegraphics[width=1\textwidth]{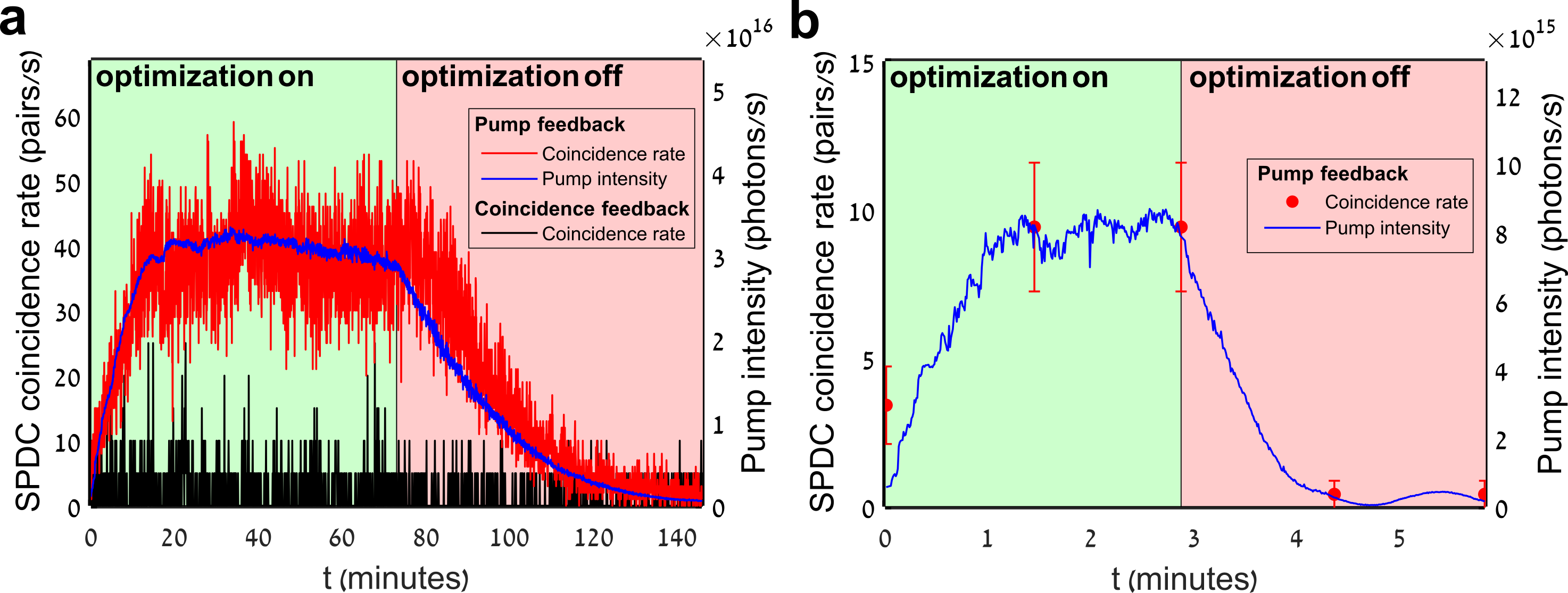}
\caption{\label{fig:3}  Real-time shaping. A diffuser is placed on a moving stage, creating a time dependent speckle pattern at the far-field. By using the intensity of the pump beam as feedback, real-time optimization is obtained for both the pump beam (blue) and the entangled pairs (red). When the optimization is turned off, degradation of both signals occurs due to the movement of the diffuser. In (a), the diffuser speed is relatively slow, so that the correspondence between the pump beam and the coincidence signal could be observed.  Even in this case, real-time optimization is not possible when the coincidence signal is used for the feedback  (black), due to its inherently low signal-to-noise ratio.  In (b), we significantly increase the speed of the diffuser, showing the strength of our method of using the pump beam for real-time optimization. Here, the optimization speed is limited by the SLM response time and the feedback electronics (100 ms), yet orders of magnitude faster optimization rates can be achieved using deformable mirror devices and fast electronics.}
\end{figure}

One of the main challenges in adopting wavefront shaping to real-world applications is scattering by a varying medium,  for example in communication through turbulent atmosphere\cite{jin2019genuine,Liao2017,paterson2005atmospheric}, as it requires fast modulation rates and sufficient signal-to-noise ratios (SNR) at short integration times. The need for fast modulation rates is common for both classical and quantum wavefront shaping. This can be solved by recent breakthroughs providing high modulation rates, ranging from a few tens of kHz using digital mirror devices (DMDs) \cite{conkey2012highDMD,Blochet2017} to 350kHz using 1D MEMS \cite{tzang2019wavefront}. The need for high SNR at short integration times, is in particular critical for quantum light, since quantum signals are too weak for real-time optimization\cite{Minozzi2013,peng2018manipulation}. This is usually solved by performing the wavefront optimization on an auxiliary bright laser, that is carefully co-aligned with the entangled photons so that it undergoes the exact same scattering\cite{Defienne2014,Defienne2016,Wolterink2016,HugoPRL}. While this approach enables fast optimization, it does not allow simultaneous transmission of the entangled photons since the auxiliary laser must have the same wavelength and polarization as the entangled photons and thus cannot be filtered out. For this reason, dynamical shaping, although crucial to many quantum technologies, has not been achieved so far for entangled photons. Since in our method the optimization is done entirely on the classical bright pump beam, the optimization rates can in principle be as fast as record-high classical wavefront shaping, making quantum wavefront shaping applicable for real-time applications. To demonstrate this feature, we emulate dynamical scattering by placing a diffuser on a moving stage, producing a time-dependent speckle pattern. First, we use a diffuser with a relatively long correlation time, for which the coincidence signal can be measured during the optimization (Fig.\ref{fig:3}a). When the optimization is turned on, both the pump (blue curve) and coincidence (red curve) signals are enhanced simultaneously, even though the diffuser is constantly moving. Although the diffuser has a relatively long correlation time, when the same optimization algorithm is used with the coincidence rate for the feedback, no enhancement is observed due to the poor SNR of the coincidence signal (black curve). In Fig.\ref{fig:3}b, we increase the speed of the moving diffuser, yielding a much shorter correlation time, which limits the number of coincidence measurements that can be performed. Nevertheless, the pump beam intensity is still enhanced by the optimization process (blue curve), causing the coincidence signal to increase as well (red points),  highlighting the strength of our use of the pump beam as feedback for real-time optimization.

\begin{figure}[H]
\centering
\includegraphics[width=0.75\textwidth]{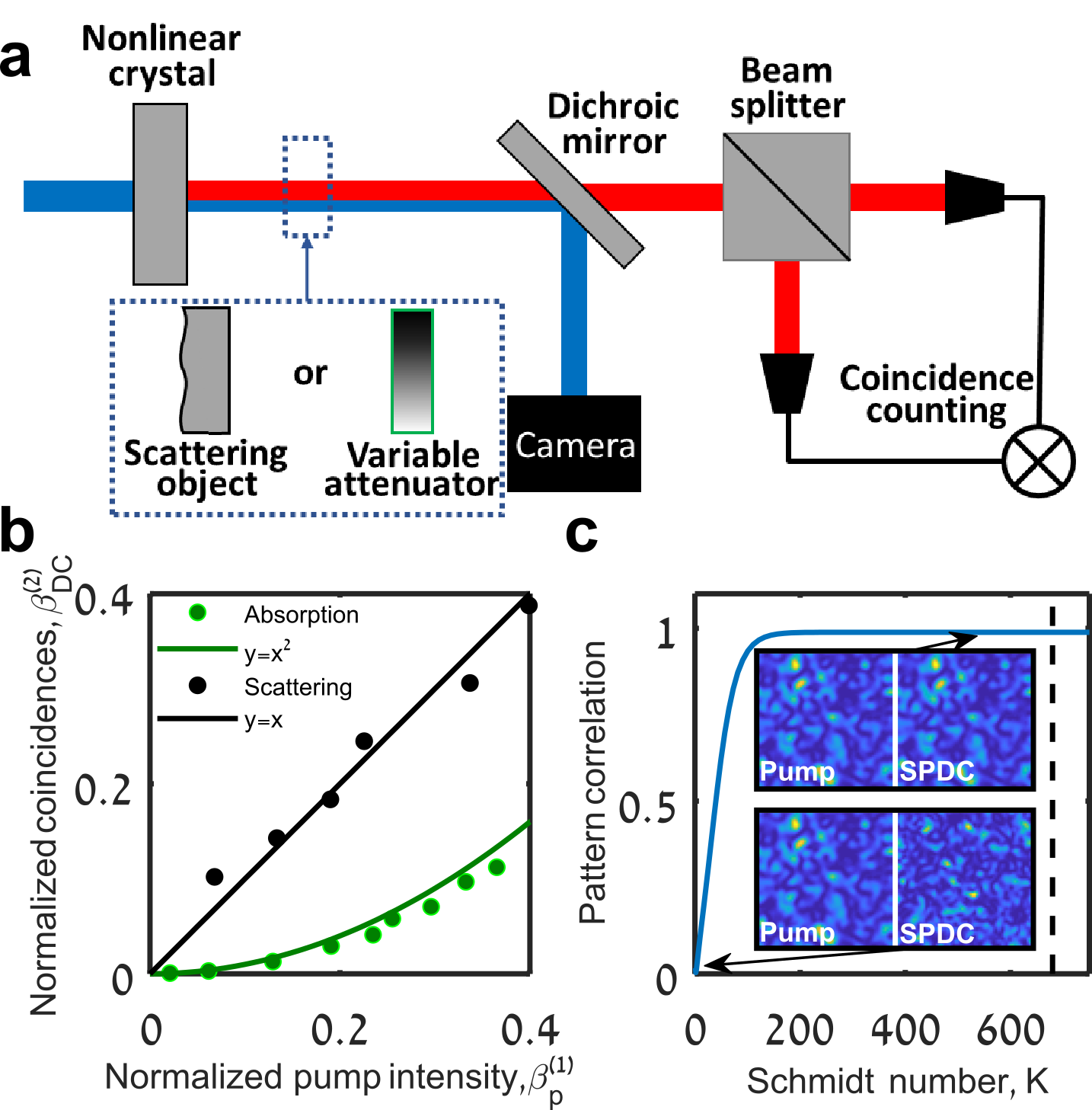}
\caption{\label{fig:4} Quantum vs. classical efficiency. (a) The two-photon coincidence rate $\beta^{(2)}_{DC}$, and the intensity of the pump beam $\beta^{(1)}_p$, both measured at the same target area and normalized by the total signal, are measured in two different configurations. In the first configuration, a linear polarizer is rotated in the optical path of both beams, to induce a variable absorption loss, yielding a clear quadratic relation between $\beta^{(2)}_{DC}$ and $\beta^{(1)}_{p}$ (b, green points). In the second configuration, the polarizer is replaced by a diffuser, and ($\beta^{(1)}_{p}$,$\beta^{(2)}_{DC}$ ) is registered during the optimization process (b, black points). Remarkably, $\beta^{(2)}_{DC}$ and $\beta^{(1)}_{p}$ exhibit a linear dependency (black curve) instead of the classically expected quadratic one (green curve). (c) The correlation coefficient between the speckle pattern of the pump beam and the coincidence pattern is presented as a function of Schmidt number. The correlation coefficient was calculated numerically under the double Gaussian approximation. In our experiment, $K\approx680$ (marked with a dashed line), yielding a high correlation between the pump and coincidence patterns which explains the linear dependency in (b).}
\end{figure}

Next, we turn to discuss the expected classical and quantum enhancements of the optimization process. Let $\beta^{(1)}_{p}$ and $\beta^{(2)}_{DC}$ be the pump intensity and entangled photons coincidence counts at the chosen target area, respectively, normalized by the total signal. We distinguish between the effects of absorption and scattering on $\beta$. In the case of absorption, the relation between the pump and coincidence signals is quadratic  $\beta^{(2)}_{DC}={\beta^{(1)}_p}^2$, as we experimentally confirm using a variable attenuator  (Fig.\ref{fig:4}a,b). This relation results from the fact that for a coincidence event, both photons must be transmitted through the absorbing media. However, for a purely scattering sample, the situation can be quite different. Although in the weak pumping regime the coincidence signal is always linear with the intensity of the pump beam\cite{gerry2005introductory}, this relation becomes more complicated when considering the signals at a target area smaller than the whole emission area, where a quadratic dependency might be expected (Supplementary). To illustrate this, we measure the two-photon signal $\beta^{(2)}_{DC}$ and the corresponding pump signal $\beta^{(1)}_{p}$ at different stages of the optimization process (Fig.\ref{fig:4}a). Remarkably, we get a clear linear dependency, $\beta^{(2)}_{DC}={\beta^{(1)}_p}$ (Fig.\ref{fig:4}b). The linear rather than quadratic dependence is a unique feature of entangled photons in the high Schmidt number regime $K\gg1$, which are scattered by a thin diffuser. To elucidate this, we numerically simulated the pump and two-photon speckle patterns formed at the far-field. The simulation was performed for two-photon states with different Schmidt numbers (corresponding to different crystal lengths), using the double-Gaussian approximation\cite{Law2004}. In Fig.\ref{fig:4}c, the calculated correlation coefficient between the two patterns is plotted versus the Schmidt number, $K$. At $K=1$, there is almost no correlation between the patterns, as expected from two different wavelengths. However, as the Schmidt number increases, the patterns become increasingly correlated, yielding almost perfect correlation in our experimental conditions at $K\approx680$. Indeed, the experimental correlation coefficient between the two patterns presented in Fig.\ref{fig:2}a,b is $0.83$.  This result completes the picture given by Eq. (\ref{2}) and Fig.\ref{fig:2}a,b, and explains the correspondence between the pump beam and entangled photon pairs in the high Schmidt number regime, that yields $\beta^{(2)}_{DC}={\beta^{(1)}_p}$. We can now easily derive the expected enhancement $\eta^{(2)}_{DC}$ of our optimization method, defined by the ratio of the coincidence rate at the target area after optimization and the average coincidence rate at the target area before optimization. As a consequence of the linear correspondence we established between the pump and two-photon signals, the coincidence enhancement  $\eta^{(2)}_{DC}$ must be exactly equal to the classical enhancement of the pump beam\cite{vellekoop2015feedback},
\begin{equation}\label{3}
\eta^{(2)}_{DC}=\eta^{(1)}_{p}=\frac{\pi}{4}(N-1)+1,
\end{equation}
where $N$ is the number of degrees of freedom used in the optimization. We therefore conclude that in our method, the efficiency of the quantum optimization is identical to the efficiency of classical wavefront shaping.

Thus far, we considered a transparent thin diffuser placed at the image plane of the nonlinear crystal. Since in this configuration the signal and idler photons pass the diffuser at the same location, it induces the same phase on both photons. The two-photon amplitude therefore accumulates the same phase as the pump amplitude, and the far-field speckle patterns of the pump beam and the entangled photons are identical. In fact, the signal and idler photons do not have to pass through the same location in the diffuser, as long as they accumulate the same phase. It is therefore enough to require that the photons pass the diffuser through the same coherence area, defined by a typical scale $d$ over which the phases induced by the diffuser become uncorrelated\cite{gatti2008three}. Pump feedback can therefore work even when the diffuser is not thin nor perfectly imaged to the crystal, as long as the thickness of the diffuser or its distance from the image plane is smaller than $d^2/\lambda$ (see Supplementary). The efficiency of the optimization will also depend on the position of the stationary detector at the far-field, since when the thickness of the diffuser is larger than its coherence length $d$, it exhibits a finite 'memory effect', i.e. a finite range of angles over which the speckle patterns remain correlated \cite{Freund1988}. The finite memory effect limits the range of target areas that can be optimized simultaneously.

\begin{figure}[H]
\centering
\includegraphics[width=1\textwidth]{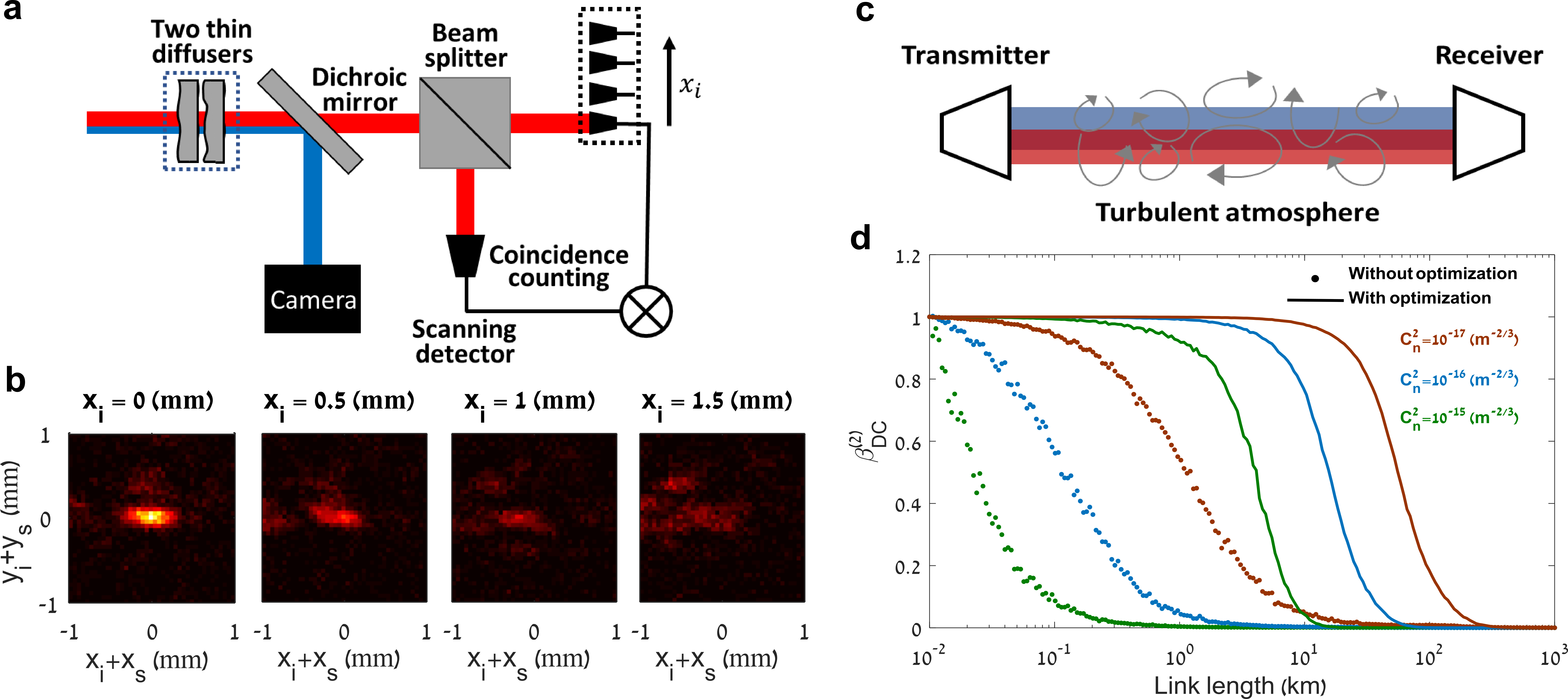}
\caption{\label{fig:5} Double diffuser configuration and atmospheric links. Two thin diffusers are placed at a distance of $3mm$ from each other to emulate a volume diffuser with a finite memory effect (a). By optimizing the pump beam, the correlations between the entangled photons at a target area on the optical axis ($x_i=0$) are enhanced. Due to the finite memory effect, when the stationary idler detector is displaced from the optical axis ($|x_i|>0$), the enhancement degrades (b). To demonstrate the applicability of pump optimization and feedback to free space optical links through turbulent atmosphere, we simulate ground to ground links with different lengths and turbulence strengths (c). The coincidence rate at the receiver end for non-optimized link (dots), decays after 10 km for weak turbulence (brown) and 100m for moderate turbulence (green). Using pump optimization (solid curves), the link length increases by two orders of magnitude, ranging from 100km for weak turbulence to 10 km for moderate turbulence. Remarkably, the increase in the link length grows with the turbulence strength (See Supplementary).}
\end{figure}

In Fig.\ref{fig:5}a,b, we demonstrate the use of pump shaping to compensate volume scattering of entangled photons by two thin diffusers separated by $z \approx 3mm$, creating an effective diffuser with thickness $z$ such that $d\ll z <d^2/\lambda$, where $d\approx 100\mu m$. This double diffuser configuration exhibits a narrow memory effect range (see Supplementary). Indeed, optimizing the pump beam makes the coincidence pattern localized (Fig.\ref{fig:5}b). When displacing the stationary idler detector, the coincidence rate at the target area decreases due to the finite memory effect associated with the double diffuser configuration.

The $d^2/\lambda$ limitation on the thickness of the diffuser restricts the range of practical applications that can benefit from pump optimization. Nevertheless, for free-space quantum links through turbulent atmosphere, pump optimization can dramatically improve the efficiency of the link, since the transverse coherence length of the atmosphere $d$ is many orders of magnitude larger than the optical wavelength $\lambda$ (Fig.\ref{fig:5}c). We utilize Kolmogorov's phase screens model for turbulence, to numerically simulate the efficiency of our pump optimization method for free-space optical links (see Supplementary). Figure \ref{fig:5}d compares the coincidence rate at the target area as a function of the length of the optical link, with and without employing pump optimization. The simulation is performed for three different turbulence strengths, ranging from weak to moderate turbulence, quantified by the structure constant of the refractive index fluctuations $C_n^2$\cite{goodman2015statistical}. A two orders of magnitude improvement in the accessible link length is achieved for all turbulent conditions. Remarkably, the improvement grows with the turbulence strength (see Supplementary).  To estimate theoretically the maximal expected link length, it is useful to introduce the Rytov variance $\sigma_R^2$, which quantifies the strength of scintillation in the optical link, where $\sigma_R^2>1$ is considered moderate to strong scintillation\cite{gbur2014rytov}. Using the reasoning presented above for a volume diffuser, we found that our method will allow efficient optimization for links with moderate scintillation, of up to $\sigma_R^2 \approx 2.5$ (see Supplementary).

\section*{Conclusions}

Our approach for shaping entangled photons, by creating them with the correct spatial correlations to compensate for the scattering, has several unique advantages over directly shaping them after their creation. First, since it is based on shaping the classical pump beam without interacting with the entangled photons themselves, it does not introduce any loss to the quantum light.  Second, since the classical pump beam is used for the optimization feedback, the optimization can be as fast as shaping of bright classical light. Moreover, since we could easily separate the pump photons from the entangled photons, we were able, for the first time, to cancel scattering of entangled-photons from a dynamically moving diffuser, without interfering their continuous transmission. This is an important step towards implementing quantum wavefront shaping in real-life scenarios,  since there is no down-time for the transmission of the photons while the optimization process is performed. 

Since the pump optimization works best for scattering layers with thicknesses that are smaller than $\frac{d^2}{\lambda}$, implementation of our method in practical applications requires that the transverse coherence length of the scattering layer $d$ will be much larger than the optical wavelength $\lambda$. As quantum communications through turbulent atmosphere falls exactly in this realm, we believe the method we developed in this work can play an important role in extending the available communication distances in scenarios where atmospheric turbulence limits the communication links\cite{Liao2017,paterson2005atmospheric,pugh2016towards}.
Finally, we note that pump shaping can be used to create a two-photon source with tailored correlations and coherence properties, by shaping both the amplitude and phase of the pump beam\cite{defienne2018partialcoher,zhang2018boydpumpcoher}, opening the door for exploiting the capacity of high dimensional qudits implemented in the spatial domain.

\section*{Methods}
The experimental setup is presented in Fig.\ref{fig:1}a. A $2mm$ long  type-0 PPKTP crystal is pumped by a $50mW$, $\lambda=404nm$ continuous-wave laser. The wavefront of the pump beam is shaped by a phase-only SLM, imaged on the crystal by two lenses with focal lengths $L1=200mm$ and $L2=100mm$ respectively. Without shaping, the pump profile at the crystal plane is approximately Gaussian with a waist of $0.7mm$. Both the pump beam and the entangled photons are then imaged onto a $0.25^{\circ}$ thin diffuser by two lenses with focal lengths $L3=100mm$ and $L4=50mm$. The pump beam and the entangled photons are separated using a dichroic mirror and measured at the far-field by an CMOS camera and $100um$ multimode fibers coupled to single photon detectors, respectively. The far-field measurements are obtained after passing through a $L5=150mm$ lens. For the coincidence measurements, $10nm$ interference filters around $808nm$ are used. In the experiment presented in Fig.\ref{fig:5}a,b, the thin diffuser is replaced with a double diffuser configuration, consisting of two thin diffusers ($0.16^{\circ}$,$0.25^{\circ}$) separated by $3mm$.

\section*{Data availability}
The data that support the findings of this study are available from the corresponding
author upon reasonable request.

\bibliography{main}

\bibliographystyle{naturemag}

\section*{Acknowledgments}
The authors thank Hugo Defienne for useful discussions. Funding: This work is supported by the Zuckerman STEM Leadership Program, the Israel Science Foundation  (grant No. 1268/16) and the Israeli ministry of science.
Authors contributions: O.L. and Y.B. designed the experiment. O.L. and G.H. built the experimental setup and performed measurements. O.L. analyzed and interpreted the data. All authors contributed to the manuscript.

\section*{Supplementary materials}

\section*{Correspondence between the pump beam and the entangled photon pairs}
As discussed in the main text, in the high spatial entanglement regime ($K\gg1$) there is a remarkable correspondence between the pump intensity distribution and the entangled photons coincidence pattern at the far-field. In this section, we demonstrate this correspondence in a simpler setting in which the diffuser in Fig.\ref{1}a is replaced by the phase mask $\phi(x,y)=sign(y)\frac{\pi}{2}$, for the wavelength of the entangled photons. Since the pump wavelength is half the wavelength of the entangled photons, the phase mask acts as a $2\pi$-step and does not affect its wavefront (Fig.\ref{fig:S1} insets). For the entangled pair the result is less obvious, since each individual photon will be affected by the induced $\pi$-step (Fig.\ref{fig:S1}a). However, when considering the pair of entangled photons, due to the high spatial entanglement, both photons hit the phase mask at the same location, thus accumulating twice the single photon phase. Therefore, although diffracting each individual photon, the two-photon wavefront is not affected by the $\pi$-step and the coincidence pattern does not change (Fig.\ref{fig:S1}b,c). In the next section, this surprising correspondence is extended to a general thin diffuser, under moderate dispersion, allowing our unique use of the pump beam for the optimization feedback.
\renewcommand{\thefigure}{S\arabic{figure}}
\setcounter{figure}{0}
\def\theequation{S\arabic{equation}}
\setcounter{equation}{0}

\begin{figure}[H]
\centering
\includegraphics[width=0.7\textwidth]{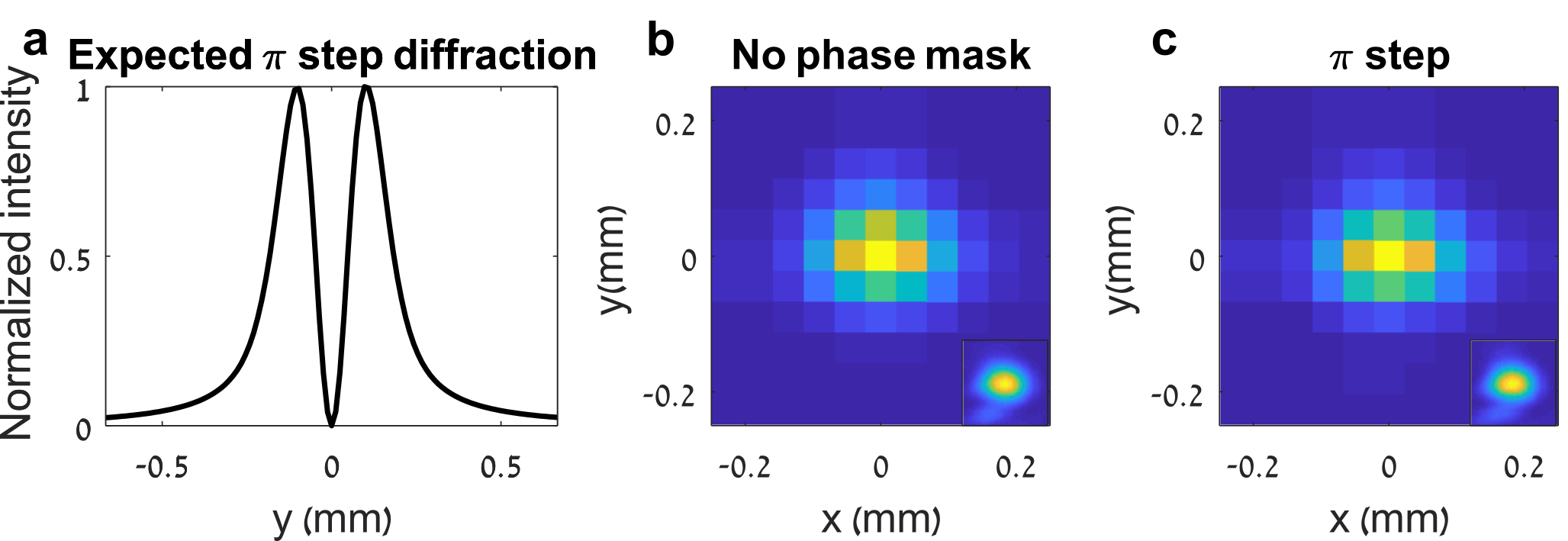}
\caption{\label{fig:S1} Pump beam and entangled photons correspondence. The diffuser in the experimental setup was replaced by an SLM, applying a $\pi$-step for the wavelength of the entangled photons (a). When no phase mask is applied (b), both the pump beam (inset) and coincidence pattern exhibit a peak with a width determined by the angular spectrum of the pump beam. When applying the $\pi$-step (c), both the pump beam and surprisingly the coincidence pattern remain unchanged.}
\end{figure}

\section*{Scattering of entangled pairs}
In the thin crystal regime, the quantum state of the entangled photons is given by  $\\|{\psi}\rangle=\int d \mathbf{q_s}d\mathbf{q_s}v(\mathbf{q_s+q_i})a^\dagger(\mathbf{q_s})a^\dagger(\mathbf{q_i})|0\rangle$, where $v(\mathbf{q})$ is the angular spectrum of the pump beam, $a^\dagger(\mathbf{q})$ is the creation operator of a photon with a transverse momentum $\mathbf{q}$  and $|0\rangle$ is the vacuum state. For simplicity, we assume here that the photons are created with the same frequency (corresponding to degenerate SPDC with narrow-band interference filters), and the same polarization (corresponding to type-0 SPDC process). The effect of a thin diffuser placed right after the crystal, can be modeled by a unitary transformation on the creation operator $a^\dagger(\mathbf{q}) \rightarrow\int d \bm{\rho} a^\dagger(\bm{\rho})A_d(\bm{\rho})\exp(i\bm{\rho}\cdot\mathbf{q})$, where $A_d(\bm{\rho})$ is the amplitude transfer function of the thin diffuser and $\bm{\rho} $ is the transverse spatial coordinate. By substituting the above relation into the quantum state we get that:
\begin{equation}\label{S1}
|{\psi}\rangle=\int d \bm{\rho} W(\bm{\rho})A_d(\bm{\rho})^2a^\dagger(\bm{\rho})^2|0\rangle
\end{equation}
where $W(\bm{\rho})$ is the spatial profile of the pump beam at the crystal plane. This representation of the quantum state stresses that, in the thin crystal regime, each pair of photons are created at the same spatial location.
Substituting the quantum state into the expression for the coincidence pattern, $C(\mathbf{q_s},\mathbf{q_i})=\left|\langle{0}|a(\mathbf{q_i})a(\mathbf{q_s})|\psi\rangle\right|^2$ , yields Eq. (\ref{2}):
\begin{equation}\label{S2}
C(\mathbf{q_s,q_i})\propto \left|\int d\bm{\rho}W(\bm{\rho})A_d^2(\bm{\rho})\exp(-i\bm{\rho}\cdot(\mathbf{q_s+q_i}))\right|^2
\end{equation}
Neglecting loss and material dispersion, we can model the diffuser as a random phase mask of the form $A_d(\bm{\rho})=\exp(i\phi(\bm{\rho}))$. In this case, the scattering of the two entangled photons is determined by an effective diffuser, given by $A_d^2(\bm{\rho})=\exp(i2\phi(\bm{\rho}))$, which corresponds to the phase accumulated by the pump beam. Thus, the created two-photon speckle pattern is identical to that of the classical pump beam, up to a scaling factor of two due to the different wavelength.

In the derivation, we assumed that the crystal is thin enough so that the phase matching condition can be neglected. To quantitatively determine the limitations on the crystal’s thickness, we consider a thin diffuser with a scattering angle $\theta\sim \lambda/d$, where $d$ is the transverse coherence length of the diffuser, i.e. the length scale over which the phase imposed by the diffuser is correlated\cite{gatti2008three}. For our method to work, we need to fulfill two conditions. 
First, both signal and idler photons must accumulate the same phase when passing through the diffuser. Therefore, we require that the spatial correlation length $d_c$ between the entangled photons is smaller than the transverse coherence length of the diffuser $d$, i.e. $d_c<d$. Since for SPDC light $d_c \propto \sqrt{\lambda L}$\cite{Walborn2010}, we get that this condition is met for $L<d^2/\lambda$, where L is the thickness of the nonlinear crystal.
Second, the angular spectrum of the shaped pump beam needs to be transferred to the quantum state of the entangled photons. This implies that the angular spread of the SPDC $\phi_{SPDC}$ has to be wider than the angular spread of the pump beam. Since the pump beam is shaped in order to compensate the scattering induced by the diffuser, the width of its angular spectrum, given in terms of angle, is $\theta$. We therefore require that $\theta<\phi_{SPDC}$. Since the angular spread of SPDC is given by $\phi_{SPDC}\propto\sqrt{\lambda/L}$\cite{Walborn2010}, we conclude that $L\theta<d$, or equivalently $L<d^2/\lambda$. Interestingly, this is the same requirement we obtained by the first condition. 

\section*{Lossy diffuser}
\begin{figure}[H]
\centering
\includegraphics[width=0.8\textwidth]{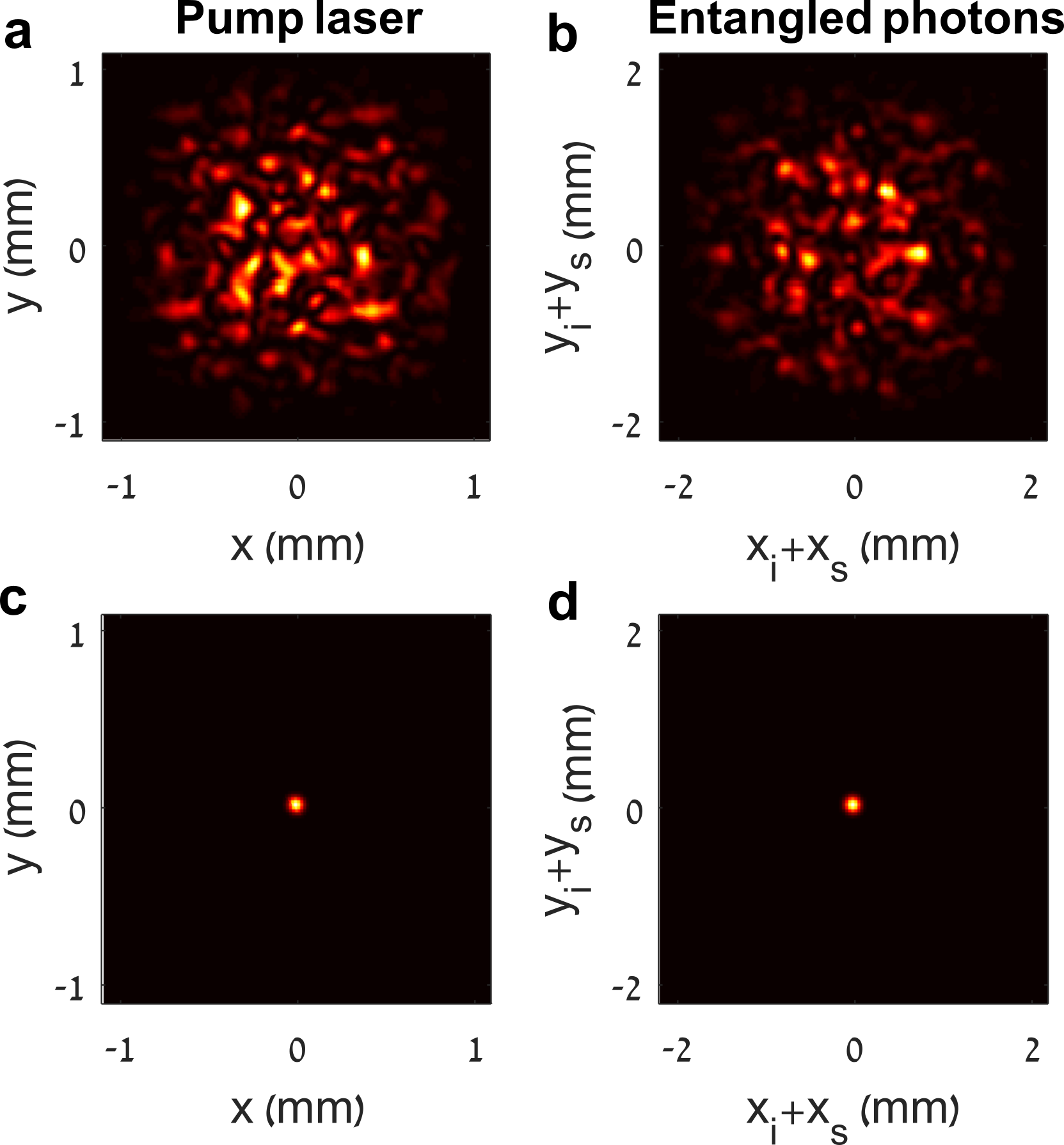}
\caption{\label{fig:S6}  Numerical simulation of a lossy diffuser. In the presence of loss, the speckle pattern of the pump beam (a) and the entangled photons (b) are not identical. Nevertheless, optimizing the phase of the pump beam (c) simultaneously optimize the coincidence signal as well (d). In this case, the effect of loss was simulated by imposing random transmission coefficients, uniformly distributed between $0$ and $1$ (see text).}
\end{figure}

In this section, we consider the performance of our pump shaping optimization method in the presence of loss. We model the lossy diffuser as an amplitude transfer function of the form $A(\bm{\rho})=t(\bm{\rho}) \exp(i\phi(\bm{\rho}))$, where $\phi(\bm{\rho})$ and $0\le t(\bm{\rho})\le 1$ are the phase and amplitude modulations imposed by the diffuser, respectively.
In this case, the coincidence pattern takes the form (Eq. (\ref{2}))

\begin{equation}\label{S7}
C(\mathbf{q_s},\mathbf{q_i})\propto \left|\int d\bm{\rho} W(\bm{\rho})t^2(\bm{\rho})\exp(i2\phi(\bm{\rho}))\exp(-i\bm{\rho}(\mathbf{q_s}+\mathbf{q_i}))\right|^2
\end{equation}
Similarly, for the classical pump beam, the intensity profile at the far-field is given by
\begin{equation}\label{S8}
I(\mathbf{q})= \left|\int d\bm{\rho} W(\bm{\rho})t(\bm{\rho})\exp(i2\phi(\bm{\rho}))\exp(-i\bm{\rho}\mathbf{q})\right|^2
\end{equation}
Inspection of the above equations shows that even in the presence of significant loss ($t^2(\bm{\rho})< 1$), since by definition both $t(\bm{\rho})$ and $t^2(\bm{\rho})$ are real and positive functions, the optimized phase correction for the pump beam, $\exp(-i2\phi(\bm{\rho}))$, will simultaneously optimize the phase of the two-photon amplitude. Remarkably, this means that in the presence of loss, the far-field speckle patterns of the pump beam and the entangled photons may not be the same, yet pump optimization will still optimize the coincidence signal, as demonstrated in Fig.\ref{fig:S6}.

One might wonder what will be the performance of phase-only optimization in the presence of significant amplitude modulations. In the common case of imperfect optimization, which can result for example from a limited number of degrees of freedom used in the optimization, the enhancement achieved by phase-only correction is lower only by a factor of $\pi/4$ compared to phase and amplitude correction\cite{vellekoop2015feedback}. In the other limit, where all phases are perfectly corrected, the residual amplitude modulations will be the limiting factor for the efficiency. However, in contrast to phase aberrations, amplitude modulations by themselves cannot severely degrade the signal, as all fields interfere constructively at the far-field target area (Fig.\ref{fig:S6}). To quantify the upper bound for the efficiency set by the residual amplitude modulations, we assume the random transmission coefficients $t_i$ are distributed with a given mean $\mu$ and standard deviation $\sigma$. In this case, the fraction of transmitted signal at the target area is given by $\langle(\sum_{i=1}^{n}t_i)^2\rangle / \langle n \sum_{i=1}^n t_i^2\rangle$, where $n$ is the number of random transmission coefficients. For large $n$, this upper bound for the efficiency approaches $(1+ \sigma ^2/\mu^2 )^{-1}$.

To demonstrate the performance of our pump shaping method in the presence of loss, we numerically simulate a segmented lossy diffuser, where each segment is assigned with random phase taken from a uniform distribution between 0 to $2\pi$, $\phi\sim unif(0,2\pi)$ and a random amplitude taken from a uniform distribution $t\sim unif(1-s,1)$, where $0\le s \le1$ is a parameter controlling the amount of loss in the simulated diffuser. When considering scattering due to amplitude modulations, the average transmission value is not by itself important, but rather the strength of the modulations around the mean. Therefore, aside from its simplicity, we chose our model such that it demonstrates strong amplitude fluctuations for $s=1$ (transmission coefficients are randomly distributed between zero and one) and no loss nor fluctuations for $s=0$ (the transmission is one).

\begin{figure}[H]
\centering
\includegraphics[width=0.8\textwidth]{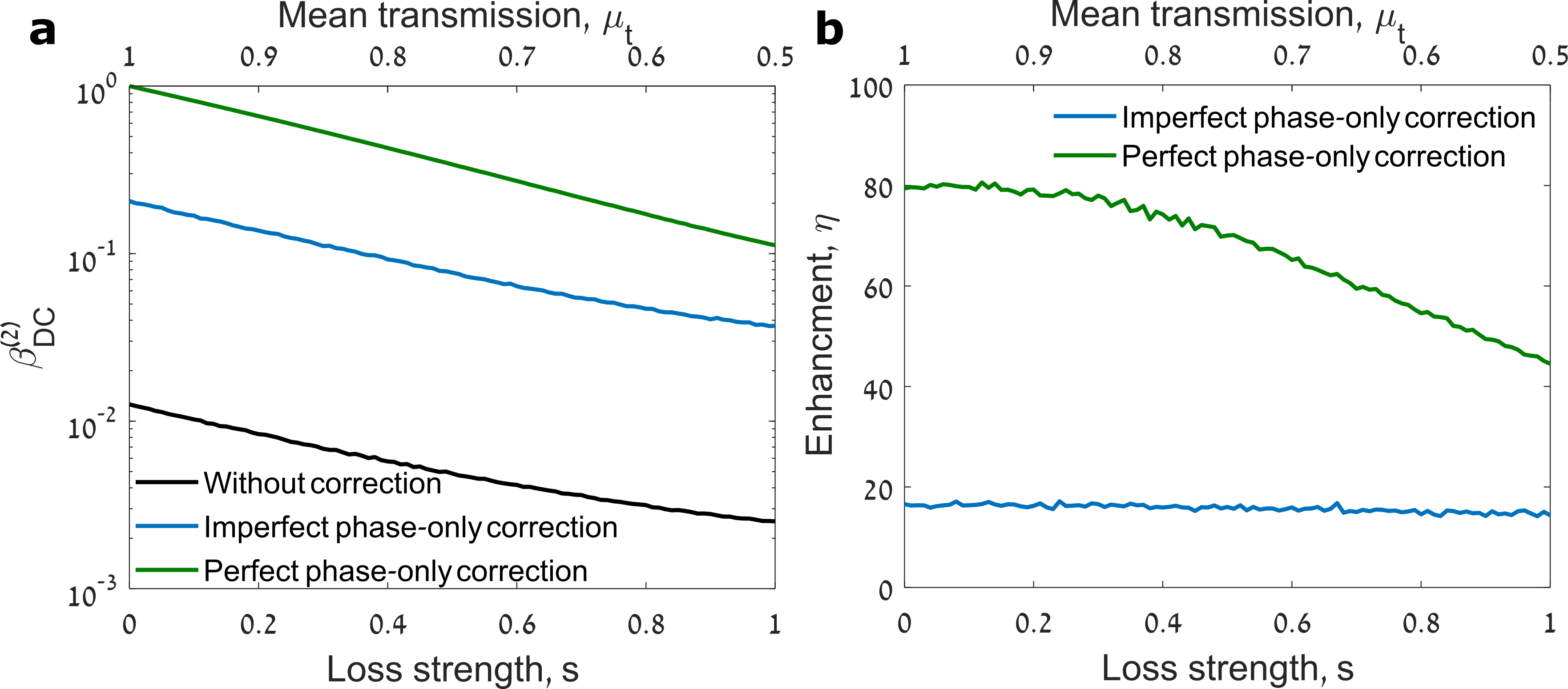}
\caption{\label{fig:S5}  Lossy diffuser. (a) The normalized coincidence signal at the target area, $\beta_{DC}^{(2)}$, as a function of the loss strength parameter $s$ chosen in the simulation. For $s=1$, the transmission coefficients follow $t\sim unif(0,1)$. (b) The enhancement achieved by the pump optimization method as a function of the loss strength, for perfect and imperfect phase-only correction. Imperfect correction was simulated by limiting the number of degrees of freedom used in the correction. In all numerical simulations, as in the experiments, the feedback for the optimization is always based on the intensity of the pump beam at the far-field target area.}
\end{figure}

From Fig.\ref{fig:S5}a, one can see that by optimizing the phase of the pump beam, the coincidence counts of the entangled photons are enhanced as well, for both perfect and imperfect phase-only corrections, even in the presence of significant loss, $t\sim unif(0,1)$. The case of imperfect phase only correction was simulated by limiting the amount of degrees of freedom used in the optimization. In Fig.\ref{fig:S5}b, we consider the enhancement $\eta$ achieved by optimizing the pump beam, as a function of the loss strength. In the case of perfect phase-only correction, the residual amplitude modulations cause the enhancement to decrease relatively to the lossless case. The decrease in the enhancement sets an upper bound for the efficiency of the optimization in the perfect phase correction case, which depends on the statistics of the random transmission coefficients. For our model with s=1, the maximal efficiency that could be achieved is bounded by $\sim 0.5$. In the case of imperfect phase-only optimization, since significant uncorrected phase aberrations are still present, the effect of the amplitude modulations is less pronounced, yielding almost constant enhancement.
Therefore, we can conclude that in the case of a lossy diffuser, one can still use phase-only pump shaping to compensate the scattering of the entangled photons, which will significantly enhance the signal at the target area.

\section*{Optimization efficiency}
In this section, we give an intuitive explanation for the surprising linear correspondence between the classical pump beam and the entangled photons (Fig.\ref{fig:4}).  When an entangled pair passes through the diffuser, both of the photons are scattered, each having a certain probability to arrive at the target area, which is much smaller than the total emission area. Following this consideration, since for a coincidence event both photons must arrive at the target area, one might naively expect a quadratic relation between the efficiency of the coincidence signal and that of a laser beam with the same optical properties. However, this is not true when considering spatially entangled photons in a high dimensional Hilbert space. In this case, the full quantum state, consisting of a large superposition of entangled pairs, must be taken into account. To explain the effect of this large superposition, we define $A,B$ as the events where the signal and idler photons arrive at the target area, respectively. Therefore, requiring that for a coincidence event both photons arrive at the target area, the efficiency of the coincidence signal is given by, $P(A\cap{B})=P(B)P(A|B)$ , where $\cap$ is the logical 'and', $P(B)$ is the ratio between the rates of the event $B$ with and without the diffuser, and $A|B$ is the conditional event of $A$ happening, given $B$.  Since the quantum state consists of a large number of spatial modes, the contrast of the single photon speckle pattern approaches zero in the high entanglement regime (Fig.\ref{fig:2}c,f). Hence, the rate of idler photons arriving at the target position does not change due to the scattering, yielding $P(B)=1$. The second term, $P(A|B)$, is determined by the spatial distribution to which the signal photon collapses after passing through the diffuser, given that the idler photon was measured at the target area. However, according to Eq. (\ref{2}) and Fig.\ref{fig:2}a,b, this spatial distribution corresponds to that of the classical pump beam, yielding the surprising linear correspondence between the pump beam intensity and the entangled photons correlations.

\section*{Volume diffusers}
\begin{figure}[H]
\centering
\includegraphics[width=0.7\textwidth]{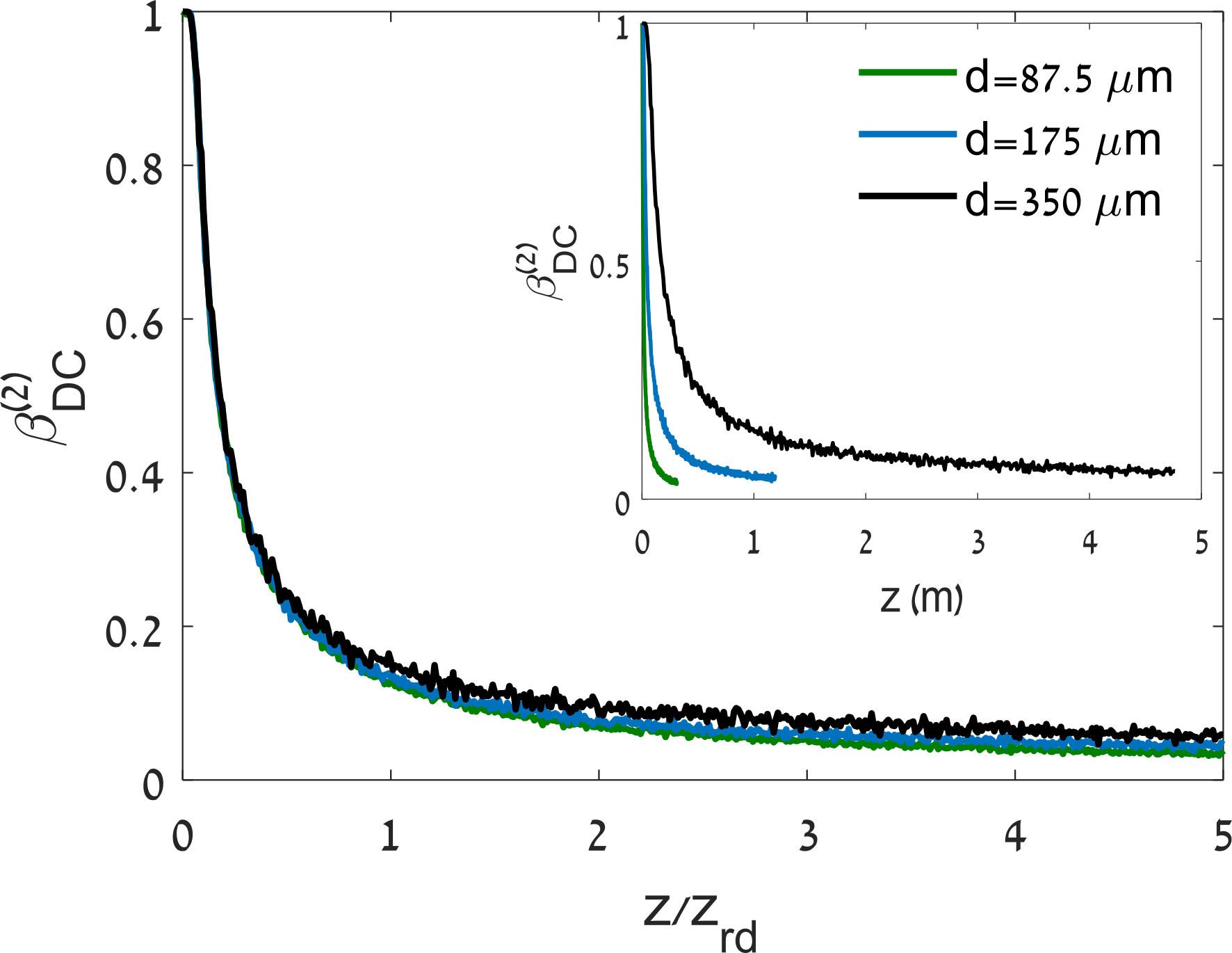}
\caption{\label{fig:S2} Volume scattering. The fraction of signal at the target area, which is our definition for the efficiency of the optimization, is simulated for a double diffuser configuration as a function of the distance between the diffusers. It is clearly seen that the efficiency of the optimization depends on the typical length scale $z_{rd}=\pi d^2/\lambda$.}
\end{figure}

As described in the previous sections, our method relies on the fact that when the signal and idler photons pass through the diffuser at the same location, their joint quantum state accumulates twice the single photon phase, which is thus equal to the phase accumulated by the pump beam.
In fact, the photons do not have to pass the diffuser at exactly the same location to accumulate the same phase. It is enough to demand that the distance between them is smaller than the coherence length of the diffuser $d$. Thus, for a diffuser with a scattering angle $\theta$, the performance of our pump shaping method will depend on the dimensionless parameter $d/z\theta$, where $z$ is thickness of the diffuser. One might note that we use the scattering angle $\theta$ in the above expression and not the angular spread of the entangled photons, $\phi_{SPDC}$. This is because even if $\theta\ll\phi_{SPDC}$ for a given target area at the far-field, only photons emitted at a range of angles $\theta$ will be scrambled by the diffuser and contribute to the signal at the target point. As for classical light, the range of far-field target positions for which the optimization works will be limited by the memory effect\cite{Freund1988}. Therefore, within the memory effect, the feedback provided by the pump beam will work up to a thickness of $z\propto d/\theta \sim d^2/\lambda$. To show this, we performed a numerical simulation of volume diffusers, emulated by two thin diffusers located at variable distance from each other (Fig.\ref{fig:S2}). Indeed, for all simulated diffuser strengths, the same scaling is observed, showing that the optimization works up to the relevant $d^2/\lambda$ scale.

The same considerations could be applied in the case of a non-imaged diffuser as well, which can be viewed as a simplified case of volume scattering in which propagation occurs prior to scattering. Therefore, by identifying the thickness $z$ from the previous analysis to be the distance of the diffuser from the image plane of the crystal, we get that the same dimensionless parameter $d^2/\lambda z$ will govern the efficiency of the optimization.
Thus, optimizing the pump beam could be useful in cases of volume scattering and non-imaged diffusers as well, within the memory effect, as long as the propagation distance is smaller than the typical propagation length $d^2/\lambda$ determined by the diffuser properties.

\section*{Diffuser characterization}
\begin{figure}[H]
\centering
\includegraphics[width=0.7\textwidth]{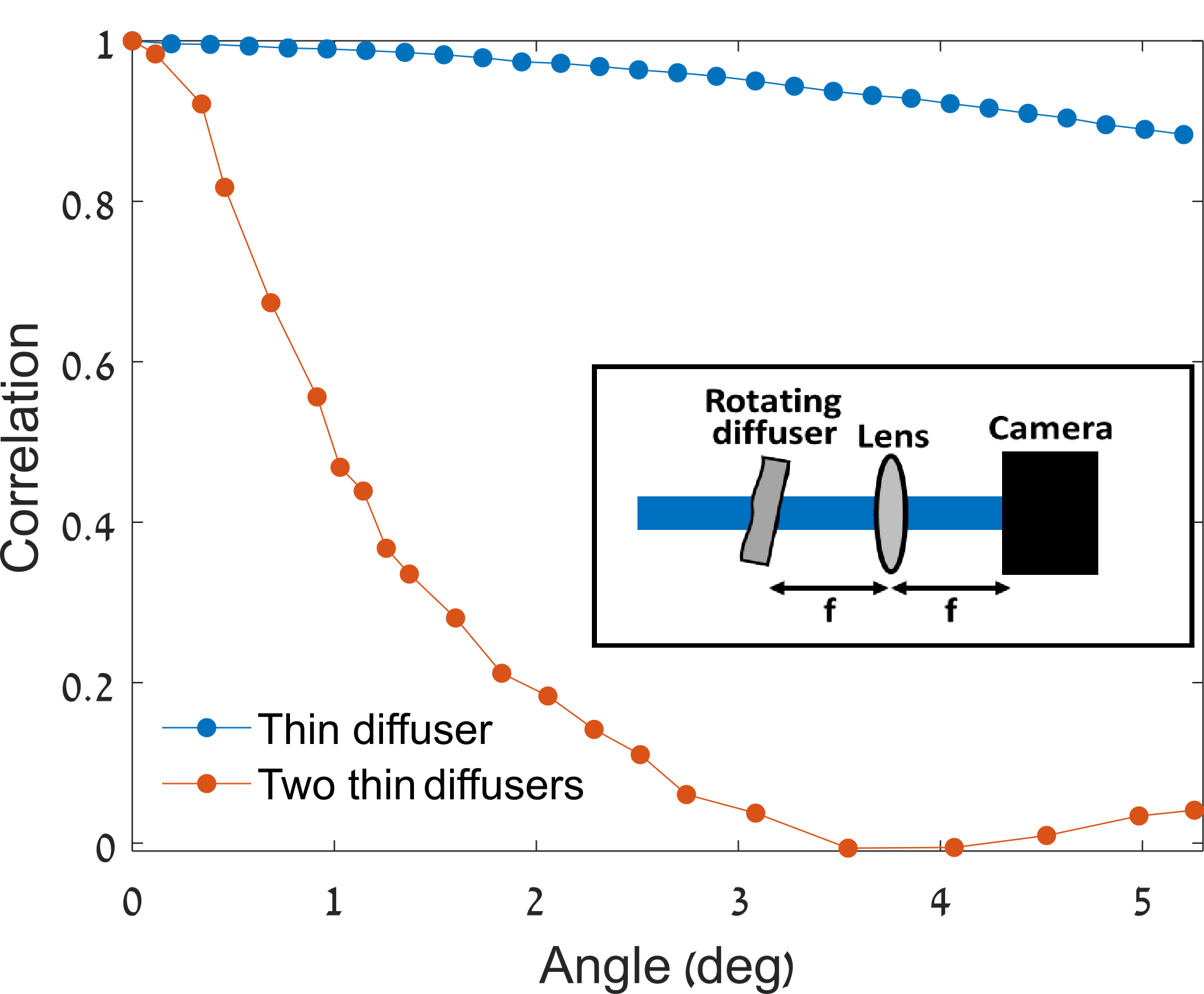}
\caption{\label{fig:S3} The memory effect of the thin diffuser (blue) and the double diffuser configuration emulating a volume diffuser (orange) was measured by rotating the diffuser and record the far-field angle dependent speckle pattern (inset). The correlation between the speckle patterns decrease significantly faster with angle for the double diffuser configuration.}
\end{figure}

In this section, we measure the memory effect for both diffusers used in the experiments. The diffusers were placed on a rotating stage, allowing us to measure the far-field speckle pattern as a function of the diffuser’s angle (Fig.\ref{fig:S3} inset). The memory effect of the double diffuser configuration, consisting of two thin diffusers separated by $3mm$, is significantly narrower than the one obtained with a single thin diffuser (Fig.\ref{fig:S3}). The transverse coherence length of the double diffuser, $d\sim \lambda/\theta$, is approximately $100\mu m$, satisfying $d\ll z$.

\section*{Propagation through turbulent atmosphere}
In this section, we explain in detail the simulation presented in Fig.\ref{fig:5}d. First, we begin by a brief review on light scattering in the atmosphere. The refractive index of air, including the effect of dispersion, is given by $n(P,T,\lambda)=1+77.6(1+7.52*(10)^{-3}/\lambda^2 ) \frac{P}{T}*(10)^{-6}$, where $\lambda$ is the wavelength in microns, $P$ is the atmospheric pressure in millibars and $T$ is the temperature in Kelvins\cite{goodman2015statistical}. The refractive index variations in the atmosphere of earth are dominated by temperature fluctuations, resulting from turbulent air flow. These refractive index fluctuations can be described by their power spectral density (PSD) function, which, according to Kolmogorov's theory of turbulence, is given by $\Phi_n (\mathbf{k})=0.033C_n^2 k^{-11/3}$, where $\mathbf{k}=(k_x,k_y,k_z)$ is the spatial frequencies vector and $C_n^2$ is the structure constant of the refractive index fluctuations, which quantifies the local turbulence strength\cite{goodman2015statistical}. To obtain a more accurate model, two corrections are often added to the PSD function, yielding the von-Karman spectrum

\begin{equation}\label{S5}
\Phi_n (k_x,k_y,k_z )=0.033C_n^2(k_x^2+k_y^2+k_z^2+k_o^2)^{-11/6} \exp(-(k_x^2+k_y^2+k_z^2)/k_m^2)
\end{equation}

where $k_o=2\pi/l_o$, $l_o$ is the so-called outer scale of the turbulence, which is usually in the 1-100m range, $k_m=5.32/l_i$  and $l_i$ is the inner turbulence scale, which has a typical value of a few millimeters\cite{goodman2015statistical}. 
To simulate the propagation of light through turbulent atmosphere, the von-Karman PSD function is transferred into a two dimensional form, representing a phase screen rather than the three dimensional refractive index fluctuations\cite{rickenstorff2016programmable}
\begin{equation}\label{S6}
\Phi_\phi (k_x,k_y)=0.49r_0^{-5/3} (k_x^2+k_y^2+k_o^2 )^{-11/6} \exp⁡(-(k_x^2+k_y^2)/k_m^2)
\end{equation}

where $r_0=(0.4229(2\pi/\lambda)^2 z C_n^2 )^{-3/5}$ is the atmosphere coherence width, sometimes called the Fried parameter\cite{rickenstorff2016programmable,fried1965statistics}, and $z$ is the atmospheric link length. For simplicity we assumed in the expression for $r_0$ a constant $C_n^2$ profile. 
To simulate propagation through thick atmosphere, we utilize a well-known two phase screen model\cite{rickenstorff2016programmable,rodenburg2014simulating}. The screens are positioned in an equally spaced configuration, and the  Fried parameter of each screen is chosen such that both the Fried parameter and scintillation effects of the entire link are correctly accounted for\cite{rickenstorff2016programmable}. Indeed, we found the effect of simulating more than two phase screens to be negligible. We simulate the random phase screens from the von-Karman PSD function using the standard inverse-Fourier transform method, with additional ten subharmonics to account for the low spatial frequency components\cite{schmidt2010numerical,lane1992simulation}. We use typical values of $l_o=10m$ and $l_i=5mm$ for the outer and inner turbulence scales. In all simulations, the pump is taken to be a Gaussian beam with a one-meter waist, chosen to ensure significant scattering by the atmosphere. We note that in all simulations the dispersion of the atmosphere was taken into account, via the above expression for $n(P,T,\lambda)$. Nevertheless, we find that the effect of dispersion is negligible.

\begin{figure}[H]
\centering
\includegraphics[width=0.7\textwidth]{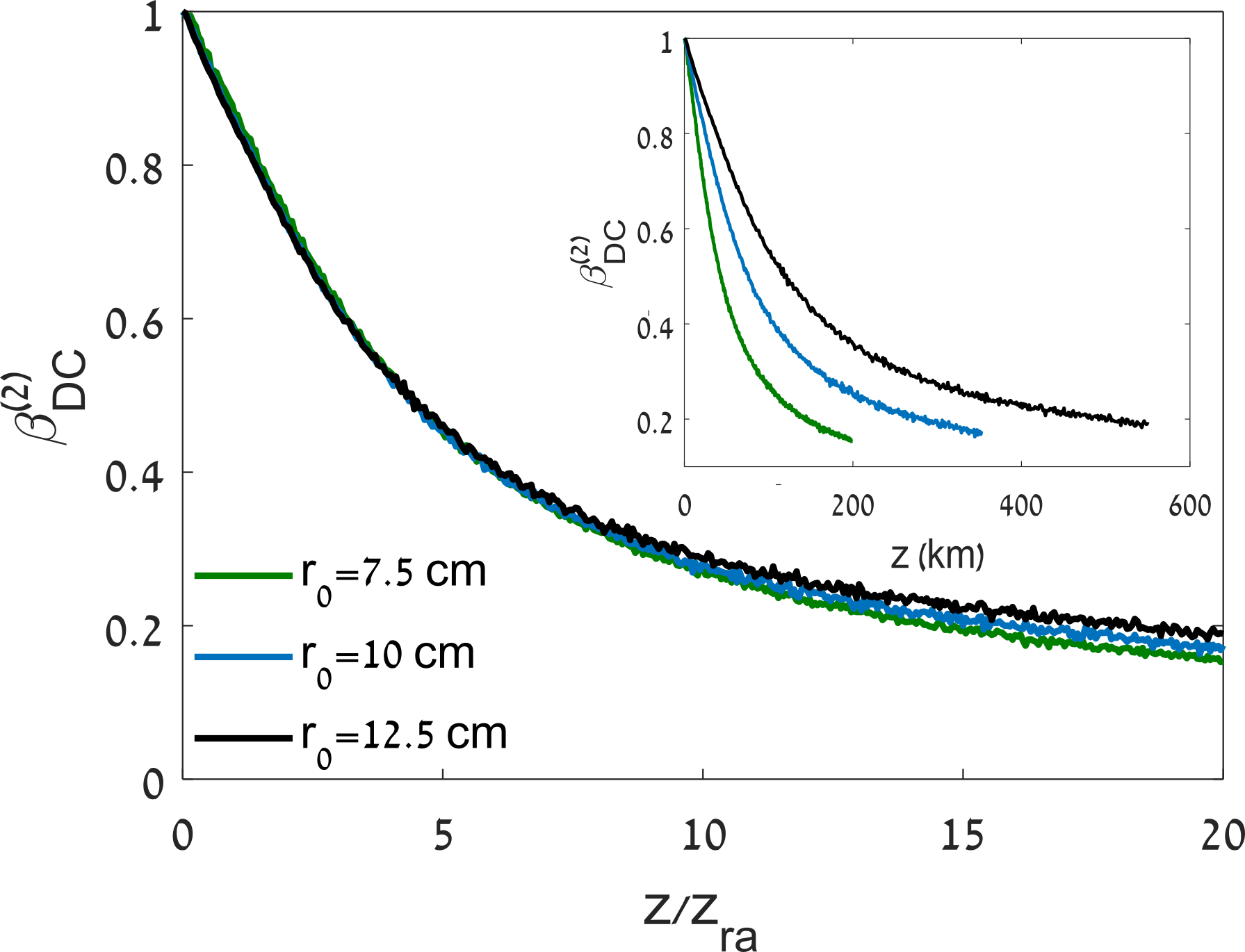}
\caption{\label{fig:S4} Thick atmosphere scaling. The efficiency $\beta_{DC}^{(2)}$ is plotted as a function of the link length for different values of $r_0$ (inset). The degradation in efficiency is determined by the typical length scale $z_{ra}=\pi\rho_0^2/\lambda$, where $\rho_0=\frac{r_0}{2.1}$.}
\end{figure}

As in the case of a volume diffuser discussed earlier, the efficiency of our method will depend on the ratio $z/z_{ra}$ , where $z_{ra}=(\pi\rho_0^2)/\lambda$, $\lambda$ is the wavelength of the pump beam and $\rho_0=r_0/2.1$ is the atmosphere coherence radius. The coherence radius $\rho_0$ is the atmospheric equivalent of the diffuser’s transverse coherence length $d$ discussed in the previous section. To show this, we simulated atmospheric links with different values of $r_0$, and calculated $\beta_{DC}^{(2)}$ as a function of the link length (Fig.\ref{fig:S4}). We can therefore derive a condition for the applicability of pump shaping to free space atmospheric links, by demanding that $z<7z_{ra}$, where the numerical prefactor was taken for a 1/e decay. It is convenient to rewrite this condition using the Rytov variance, $\sigma_R^2=1.23(2\pi/\lambda)^{7/6} C_n^2 z^{11/6}$, which quantifies the scintillation strength in the optical link, yielding $\sigma_R^2<2.5$. As $\sigma_R^2$ defines the transition between weak to strong turbulence\cite{gbur2014rytov}, we conclude that our pump shaping method could be used in a wide range of free space optical links, including both ground to ground and ground to satellite links with moderate turbulent conditions. In Fig.5d in the main text, three ground to ground links with different $C_n^2$ values were simulated using the method described before. In this case, $C_n^2$ was kept constant while the link length is increased, yielding stronger scattering (smaller $r_o$) and scintillation. A two orders of magnitude improvement in the link length is observed.

The enhancement of the link length due to pump optimization can be calculated based on the analysis presented above. We define $z_o$ to be the link length for which the efficiency of the quantum signal drops to half, when pump optimization is employed. Similarly, $z_{no}$ is the equivalent length in the non-optimized case. Since the efficiency of the pump optimized quantum signal depends on $(\pi\rho_0^2)/\lambda z$, we get that $z_o$ satisfies $0.5\sim (\pi \rho_0^2)/(\lambda z_o )$. For the non-optimized case, the efficiency depends on $(r_0/w)^2$, where w is the waist of the Gaussian beam\cite{andrews2004atmospheric}. Therefore, $z_{no}$ satisfies $0.5\sim (r_0/w)^2$. Combining these two results and substituting the expressions for $r_0$ and $\rho_0$, we get that the enhancement of the link length remarkably increases with the turbulence strength $C_n^2$, and is given by
\begin{equation}\label{S9}
\frac{z_o}{z_{no}} \propto \frac{w^{5/3}}{\lambda ^{15/11}}(C_n^2)^{5/11}
\end{equation}

We demonstrate this result in Fig.S7, where we simulate $z_o/z_{no}$  as a function of $(C_n^2 )^{5/11}$, showing a clear linear relation.

\begin{figure}[H]
\centering
\includegraphics[width=0.7\textwidth]{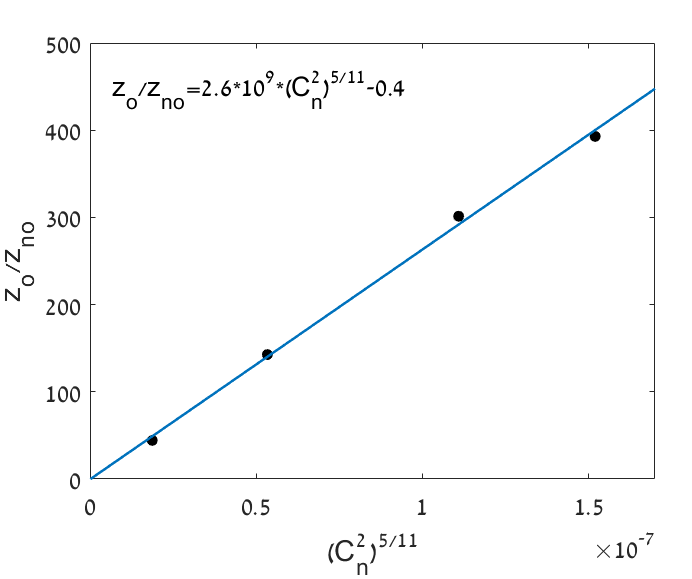}
\caption{\label{fig:S7} The ratio between the optimized and non-optimized link lengths, defined by the length for which the efficiency drops to half, is presented as a function of $(C_n^2 )^{5/11}$. Dots are obtained from a numerical simulation, and the curve is a linear fit given by the equation in the inset. The clear linear relation indicates that the improvement offered by our method remarkably increases as the turbulence becomes stronger.}
\end{figure}

\begin{figure}[H]
\centering
\includegraphics[width=0.7\textwidth]{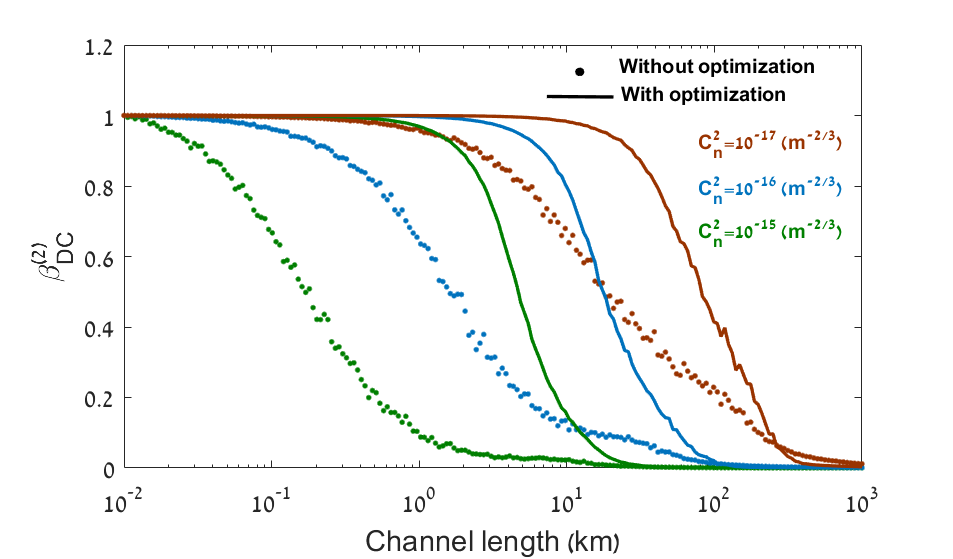}
\caption{\label{fig:S8} The simulation presented in Fig.\ref{fig:5}d is repeated using a Gaussian beam with an 15cm waist at the transmitter plane. Our optimization method demonstrates significant improvement in the link length in this case as well.}
\end{figure}

Finally, we note again that in order to ensure a significant degradation by the turbulent atmosphere, we used the extreme case of a Gaussian beam with a large, one-meter waist. To show that our approach provides a significant advantage in more common cases as well, we repeated the simulation in Fig.\ref{fig:5}d with a waist of 15cm. The results are presented in Fig.\ref{fig:S8}, showing a clear advantage provided by our approach in this case as well.

\section*{Measurement of the Schmidt number}
In order to estimate the Schmidt number $K$ in our experimental apparatus, we use the double-Gaussian approximation for the phase matching condition, yielding a two-photon distribution of the form\cite{Law2004},
\begin{equation}\label{S3}
\psi(\mathbf{q_s,q_i}) \propto exp\left(-\frac{(\mathbf{q_s+q_i})^2}{\sigma^2}\right)exp\left(-b^2(\mathbf{q_s-q_i})^2\right) 
\end{equation}
where $b^2=\frac{L}{4k}$ , $k$ is the pump wavenumber inside the crystal, and $\sigma$ is the waist of the Gaussian pump beam. 

Under the double-Gaussian approximation, the Schmidt number $K$ could be computed exactly, and is given by\cite{Law2004},
\begin{equation}\label{S4}
K=\frac{1}{4}\left(\frac{1}{b\sigma}+b\sigma\right)^2
\end{equation}
Thus, in the thin crystal or weak focusing regimes, where $b\sigma\ll1$, one could approximate the Schmidt number as $K=\frac{1}{\left(2b\sigma\right)^2}$.
This expression for the Schmidt number can be understood intuitively by noticing its relation to the number of transverse spatial modes in the system\cite{VanExter2006}. Therefore, the Schmidt number could be estimated experimentally by using the width of the far-field single and coincidence counts distributions as follows. The coincidence pattern at the far-field is given by $C(\mathbf{q_s,q_i}=0)\propto \left|\exp(-\mathbf{q_s^2}/\sigma^2)\exp(-b^2\mathbf{q_s^2})\right|^2$, where we assumed for simplicity that the stationary detector is placed such that $\mathbf{q_i}=0$. Since in the thin crystal regime $\sigma\ll1/b$, we can approximate $C(\mathbf{q_s,q_i}=0)\propto \left|\exp(-\mathbf{q_s^2}/\sigma^2)\right|^2$, and thus measure $\sigma$ from the width of the coincidence pattern. The single photon counts distribution, which is obtained by scanning one of the detectors and recording the photon rate, is given by $I(\mathbf{q})=\left| a(\mathbf{q})|\psi\rangle\right|^2\propto \int d\mathbf{q'}\left|\exp⁡(-(\mathbf{q+q'})^2/\sigma^2)\exp⁡(-b^2(\mathbf{q-q'})^2 )\right|^2$. Since $\sigma\ll1/b$, we can approximate $\mathbf{q'}\approx-\mathbf{q}$, yielding $I(\mathbf{q})\propto\left|\exp⁡(-b^2 (2\mathbf{q})^2 )\right|^2$. Thus, $b$ can be measured from the single counts distribution width. Combining these two measurements, we obtained a Schmidt number $K=680 \pm 60$ in our experiment, indicating high spatial entanglement.
\end{document}